\documentclass[twocolumn,prb,aps,showpacs]{revtex4}
\usepackage{mathrsfs}
\usepackage{graphicx}
\begin{document}
\title {Impurity effects on the quantum coherence of a few-boson system}
\author{Pei Lu}
\author{Zhi-Hai Zhang}
\author{Shiping Feng}
\author{Shi-Jie Yang\footnote{Corresponding author: yangshijie@tsinghua.org.cn}}
\affiliation{Department of Physics, Beijing Normal University, Beijing 100875, China}
\begin{abstract}
The impurity effects on the quantum coherence of a few-boson system are studied within the two-site
Hubbard model. Periodical collapses and revivals of coherence occur in the presence of either polarized
or unpolarized fermionic impurities. The relative strength $U_{BF}/U_{BB}$ of the boson-fermion
interaction versus the boson-boson interaction plays a key role in the coherence revivals. As the
average filling of the impurity increases, the coherence revivals remain nearly unaffected for
$U_{BF}/U_{BB}=z$ ($z\in Z$ is an integer) while the odd revival peaks are damped for
$U_{BF}/U_{BB}=z+1/2$, in agreement with the experimental observations. For unpolarized fermionic
impurities, the coherence revivals are irrelevant to the strength of the fermion-fermion interactions.
\end{abstract}
 \maketitle

\section{introduction}
The ultracold atoms, trapped either by a magnetic potential or an optical lattice, present an
advantageous phase of matter for the investigation of fundamental quantum
physics\cite{Paganelli,Xu,Albus,Imambekov,Messeguer,Chatterjee,Zhou}. Since the external potentials, the
dimensions, the effective interactions, as well as the atomic components, can be well-controlled and
precisely measured, the ultracold atoms have been providing versatilely and experimentally feasible
means to explore a variety of problems such as quantum simulation, information processing, strongly
correlated systems, dynamical evolution, and so on. In particular, atomic interferometers have been realized experimentally by loading Bose-Einstein condensates (BECs) into double-well potentials. Atomic Josephson
oscillations\cite{Javanainen,Jack,Zapata} and macroscopic quantum self-trapping\cite{Anker,Albiez} were
predicted and observed experimentally. The phenomena of quantum tunneling\cite{Winkler,Folling,Zollner},
the collapse and revival of quantum coherence \cite{Best,Greiner,Yang}, disorder effects
\cite{Fallani,Lye,Fort,Zhang}, and paired or counterflow superfluidity \cite{Kuklov,Yang1} are
extensively explored.

The problem of impurity embedded in a quantum environment poses another interesting
topic\cite{Zipkes,Schmid}. In a recent experiment, the authors studied the quantum dynamics of bosons in
an optical lattice with a fraction of polarized fermionic impurities\cite{Will}. The absolute strength
of the intraspecies and interspecies interactions are measured as a function of the interspecies scattering
length, tuned by means of a Feshbach resonance. The collapses and revivals of the quantum coherence
exhibit distinct features in comparison to the pure bosonic system\cite{Greiner,Will}. One of the most
prominent facts are the odd revival peaks are damped as the impurity fillings increase. When such
impurity systems are scaled down to the few-body regime, the mean-field theory becomes
invalid\cite{Will,Mulansky}. The double-well model provides the direct ways to go beyond the
Gross-Pitaevskii paradigm of gaseous BECs. It is also one of the simplest prototypes for finite lattices
and the most instructive means to study quantum dynamics.

In this work, we explore the impurity effects on the quantum coherence of a few-boson system. For
bosonic atoms comprising fermionic impurities in the optical lattices, the system can be properly
treated as a mixed ensemble of the pure bosonic system and the bosonic system with a single fermionic
impurity in the double-well model. We deal with $N$ interacting bosons with a small fraction of
fermionic impurities. Under the approximation of tight binding, the double-well is simplified as a
two-site single-band Hubbard model\cite{Jaksch,Vardi}. The boson (fermion) creation $\hat{b}_i^\dag$
($\hat{f}_i^\dag$) and annihilation $\hat{b}_i$ ($\hat{f}_i$) operators are constructed for atoms
localized in either side of the well. In the limit of strong interactions $U_{BB}/t_B\gg 1$, the system
which starts from a initial coherence state experiences periodical collapses and revivals of coherence.
The relative strength of the boson-boson interaction versus the boson-fermion interaction
$U_{BF}/U_{BB}$ plays an important role on the dynamical revival and collapse of macroscopic matter
waves. The results agree with the experimental observations in Ref.[\onlinecite{Will}]. It is discovered
that symmetrical and anti-symmetrical coherent states alternatively occur at the revival peaks. We
further investigate the effects of unpolarized fermionic impurities on the quantum coherence. The
coherence revivals are irrelevant to the interaction strength between the impurities.

The paper is organized as follows: In Sec.II, we illustrate the model and the formulism. In Secs.III and
IV we explore the effects of polarized and unpolarized fermionic impurities on the coherence,
respectively. A brief summary is included in Sec. V.

\section{model}
We consider a mixture of a pure bosonic system and a bosonic system with a single fermionic impurity in
the double-well. The usual two-site Bose-Hubbard model for the pure bosons is written as
\cite{Jaksch,Vardi,Lee1,Lee2}
\begin{equation}
\hat{H}_0=-t_B(\hat{b}_1^\dag \hat{b}_2+\hat{b}_2^\dag
\hat{b}_1)+\frac{1}{2}U_{BB}\sum_{i=1,2}\hat{n}_i(\hat{n}_i-1),\label{ham0}
\end{equation}
and the Hamiltonian including fermionic atoms should be the Bose-Fermi-Hubbard model which is written as
\cite{Jaksch,Vardi}
\begin{eqnarray}
\hat{H}_1&=&-t_B(\hat{b}_1^\dag \hat{b}_2+\hat{b}_2^\dag\hat{b}_1)+\frac{1}{2}U_{BB}\sum_{i=1,2}\hat{n}_i(\hat{n}_i-1)\nonumber\\
&&-t_F(\hat{f}_1^\dag \hat{f}_2+\hat{f}_2^\dag
\hat{f}_1)+U_{BF}\sum_{i=1,2}\hat{n}_i\hat{m}_i.\label{ham1}
\end{eqnarray}
Here $U_{BB}$ and $U_{BF}$ are the boson-boson and boson-fermion interactions, respectively. $t_B$ and
$t_F$ are the bosonic and fermionic hopping coefficients, respectively. $\hat{n}_i$ and $\hat{m}_i$
($i=1,2$) are the bosonic and fermionic number operators, respectively. Since in the single-band Hubbard
model the double occupation of polarized fermions is prohibited, the average site-filling of the
fermionic impurities is restricted to $\bar m\leq 0.5$.

The Hamiltonian (\ref{ham0}) is represented in the Fock basis set \{$|N,0\rangle, |N-1,1\rangle, \cdots,
|0,N\rangle $\}. The eigenstates can be expressed as linear combinations of the Fock bases,
$|\psi_j\rangle=\sum_{k=0}^N c_{jk}|N-k,k\rangle$ ($j=0,1,2,\ldots,N$), which correspond to the
eigenvalues $\omega_j$. The coefficients $c_{jk}$ satisfy the recursive relation\cite{Yang}
\begin{eqnarray}
&&-t_B\sqrt{(N-k)(k+1)} c_{j(k+1)}-t_B\sqrt{(N-k+1)k} c_{j(k-1)}\nonumber\\
&&+[\frac{U}{2}(N^2-2Nk-N+2k^2)-\omega_j] c_{jk}=0.
\end{eqnarray}
The temporal evolution of the state is governed by the Heisenberg's equation for a given initial state
$|\psi(0)\rangle$,
\begin{equation}
|\psi(\tau)\rangle=\sum_{j=0}^Nf_j(\tau)|\psi_j\rangle \equiv \sum_{k=0}^N g_k(\tau) |N-k,k\rangle,
\end{equation}
where $f_j(0)=\langle\psi_j|\psi(0)\rangle$ and $g_k(\tau)=\sum_{j=0}^N f_j(0) c_{jk} e^{-i\omega_j
\tau}$.

To depict the coherence degree of the system, we introduce a characteristic parameter\cite{Yang}:
\begin{equation}
\alpha_1=\frac{|\lambda_1-\lambda_2 |}{\lambda_1+\lambda_2},
\end{equation}
where $\lambda_1$ and $\lambda_2$ are the two eigenvalues of the single-particle density
$\rho_{\mu\nu}(\tau)=\langle \psi(\tau)|\hat{a}_\mu^\dag \hat{a}_\nu |\psi(\tau)\rangle$
($\mu,\nu=1,2$)(Refs.\onlinecite{Penrose,Mueller}). When $\alpha\rightarrow 1$, the system is in the coherent
(quasi-coherent) state since in this case there is only one large eigenvalue of matrix $\rho_{\mu\nu}$.
Accordingly, $\alpha\rightarrow 0$ indicates the system is in the decoherent or fragmented state because
there are two densely populated natural orbits. In the weak-interaction, strong-tunneling limit
($U_{BB}/t_B\ll1$), each atom is in a coherent superposition of the left-well and right-well states. In the
strong interactions or weak tunneling ($U_{BB}/t_B\gg 1$), the tunneling term is negligible. The
Hamiltonian is the product of the number operators for the left and right wells. The eigenstates are
products of Fock states and are referred to as decoherent states. This regime is analogous to the Mott
insulator (MI) phase in optical lattices.

On the other hand, for boson-fermion mixed system, the Hamiltonian (\ref{ham1}) is represented in the
Fock set \{$|N,0\rangle |0,1\rangle, |N-1,1\rangle |0,1\rangle, \cdots, |0,N\rangle |0,1\rangle$,
$|N,0\rangle |1,0\rangle, |N-1,1\rangle |1,0\rangle, \cdots, |0,N\rangle |1,0\rangle $\} as
\begin{widetext}
\begin{equation}
\hat{H}_2=\left(\begin{array}{cccccc}
h_{00}& \cdots &0&-t_F&\cdots&0\\
-t_B\sqrt{(N-1)}&\cdots&0&0&\cdots&0\\
\vdots&\ddots&\vdots&\vdots&\ddots&\vdots\\
0&\cdots&\frac{1}{2}U_{BB}(N-1)(N-2)&0&\cdots&-t_F\\
-t_F&\cdots&0&\frac{1}{2}U_{BB}(N-1)(N-2)&\cdots&0\\
0&\cdots&0&-t_B\sqrt{(N-1)}&\cdots&0\\
\vdots&\ddots&\vdots&\vdots&\ddots&\vdots\\
0&\cdots&-t_F&0&\cdots&h_{00}
\end{array}\right),
\end{equation}
\end{widetext}
where $h_{00}=U_{BB}(N-1)(N-2)/2+U_{BF}(N-1)$. By solving the Heisenberg's equation of motion, we can
also obtain the temporal evolution of the state. The coherence $\alpha_2$ of the bosons is defined in
the same way as for the definition of $\alpha_1$ in a pure bosonic system.

For the ensemble of mixed systems of pure bosons and bosons with polarized fermionic impurities in the
double-well, or bosons with fermionic impurities in the optical lattice, the average coherence in
accordance to the fermion-filling $\bar{m}$ is,
\begin{equation}
\alpha=(1-2\bar m)\alpha_1+2\bar m\alpha_2.\label{coherence}
\end{equation}

In the following calculations, we choose $N=10$ bosons as example and set the units of $t_B=t_F=1$.

\section{Polarized fermionic impurities}
For the quantum dynamics, we are concerned with a delocalized Bose-Fermi mixture in a shallow optical
lattice. When the system parameters are swiftly changed, the atoms collectively experience a dynamical
evolution. In our double-well model, we focus on the typical case in which the initial state is a
symmetric coherent state. For the pure bosonic system, the initial state is written as
\begin{equation}
|\psi_B(0)\rangle=(\frac{\hat{b}_1^\dag+\hat{b}_2^\dag}{\sqrt2})^N|0\rangle.
\end{equation}
For the bosonic system with a polarized fermionic impurity, the initial state is written as:
\begin{equation}
|\psi_B(0)\rangle|\psi_F(0)\rangle=(\frac{\hat{b}_1^\dag+\hat{b}_2^\dag}{\sqrt2})^N(\frac{\hat{f}_1^\dag+\hat{f}_2^\dag}{\sqrt2})|0\rangle.
\label{initial}
\end{equation}
\begin{figure}
\begin{center}

\includegraphics*[width=8.5cm]{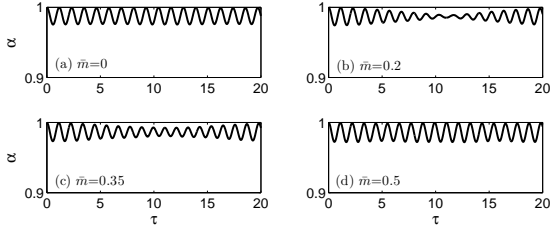}
\caption{Temporal evolution of $\alpha(\tau)$ for the $N=10$ system starting from a coherent state.
$U_{BB}=0.2$ and $U_{BF}/U_{BB}=0.5$. From (a) to (d), $\bar m=0,0.2,0.35,0.5$. For all cases,
$\alpha\rightarrow 1$.}
\end{center}
\end{figure}

When the lattice depth is swiftly changed to $U_{BB}/t_B=0.2$ which is still in the SF regime, the
coherence parameter $\alpha$ fluctuates around 1, as shown in Figs.1(a) through (d). It indicates that the
bosons preserve coherence or quasi-coherence and is nearly unaffected by the impurities.

\begin{figure}
\begin{center}
\includegraphics*[width=8.5cm]{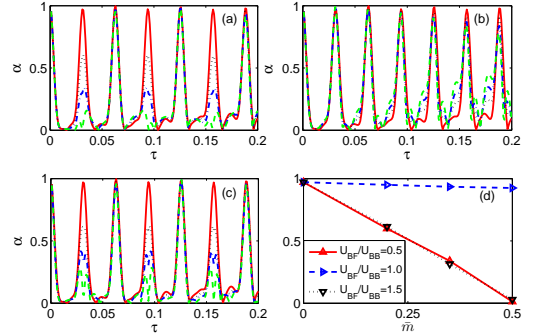}
\caption{(Color online) Temporal evolution of $\alpha(\tau)$ for the $N=10$ system starting from a
coherent state. $U_{BB}=100$. From (a) to (c), and $U_{BF}/U_{BB}=0.5,1.0,1.5$. The solid (red) line,
dotted (black) line, dash-dotted (blue) line, and dashed (green) line represent
$\bar{m}=0,\bar{m}=0.2,\bar{m}=0.35$, and $\bar{m}=0.5$, respectively. (d) The dependence of
the hight of the first revival peak on the impurity filling $\bar{m}$ is illustrated.}
\end{center}
\end{figure}

On the other hand, when the lattice depth is swiftly changed to the deep MI regimes ($U_{BB}/t_B\gg 1$),
the value of $\alpha$ oscillates between 1 and 0 with a period of $T\approx 0.031$, as shown in
Figs.2(a) through (c). It implies that the system experiences collapses and revivals of the coherence\cite{Yang}.
Notably, the odd revival peaks are gradually damped as the impurity filling increases for
$U_{BF}/U_{BB}=z+0.5$ [Figs.2(a) and (c)]. At $\bar m=0.5$, these revival peaks are ultimately destroyed.
In comparison, as $U_{BF}/U_{BB}=1$, no revival peaks are damped. The coherence revivals are not
affected by the impurities regardless of the variations of the impurity fillings [Fig.2(b)]. This
phenomenon was recently observed in the experiment\cite{Will}.

This phenomenon can be understood by the following single-site evolution model. As the system is swiftly
changed to the deep MI regimes, both the bosonic and fermionic tunneling are suppressed and the
delocalized distributions of bosons and fermions are freezed. The eigenstates at a lattice site are
given by product atom number states, containing $k$ bosons ($k=0,1,2,\cdots,N$) and $m$ fermions
($m=0,1$). The eigenenergies corresponding to the eigenstates
$|\psi_{km}\rangle=(\hat{b}_j^\dagger)^k(\hat{f}_j^\dagger)^m |0\rangle$ are
$E_{km}=U_{BB}k(k-1)/2+U_{BF}km$. The initial state can be expanded as,
\begin{equation}
|\Psi(\tau)\rangle=\frac{1}{2^{(N+1)/2}}\sum_{k=0}^N
C_N^k(\hat{b}_1^\dag)^{N-k}(\hat{b}_2^\dag)^k(\hat{f}_1^\dag+\hat{f}_2^\dag)|0\rangle,
\end{equation}
which is a linear superposition of products of the eigenstates. It evolves independently according to
\begin{widetext}
\begin{eqnarray}
|\Psi(\tau)\rangle&&=\frac{1}{2^{(N+1)/2}}\sum_{k=0}^N\{C_N^ke^{-i[U_{BB}(N-k)(N-k-1)/2+U_{BF}(N-k)]\tau}(\hat{b}_1^\dag)^{N-k}{\times}e^{-i[U_{BB}k(k-1)/2]\tau}(\hat{b}_2^\dag)^k\hat{f}_1^\dag|0\rangle\nonumber\\
&&+C_N^ke^{-i[U_{BB}(N-k)(N-k-1)/2]\tau}(\hat{b}_1^\dag)^{N-k}{\times}e^{-i[U_{BB}k(k-1)/2+U_{BF}k]\tau}(\hat{b}_2^\dag)^k\hat{f}_2^\dag|0\rangle\}\nonumber\\
&&=\frac{e^{-iU_{BB}N(N-1)\tau/2}}{2^{(N+1)/2}}\sum_{k=0}^NC_N^ke^{iU_{BB}(N-k)k\tau}(\hat{b}_1^\dag)^{N-k}(\hat{b}_2^\dag)^k[e^{-iU_{BF}(N-k)\tau}\hat{f}_1^\dag+e^{-iU_{BF}k\tau}\hat{f}_2^\dag]|0\rangle.
\label{evolution}
\end{eqnarray}
\end{widetext}

It follows that, except for a global phase factor, there is a time-dependent phase factor attached to each term in Eq.(\ref{evolution}). At time interval $T=\pi/U_{BB}$, the phases $e^{iU_{BB}(N-k)k\tau}=(-1)^k$. In the meantime, if $U_{BF}/U_{BB}=z$ ($z\in Z$) the fermionic terms also give rise to a factor of $(-1)^k$
as $U_{BF}T=z\pi$. Hence the state evolves to the initial coherent states. At other times, the
superposition from various terms in Eq.(\ref{evolution}) cancels and the coherence is destroyed. These
are the coherence collapses and revivals, with the revival period $T=\pi/U_{BB}$. On the other hand, if
$U_{BF}/U_{BB}=(z+0.5)$, the fermionic terms recover its initial state only after a time interval
$T_F=\pi/U_{BF}$. Hence the revival period should be $T=2\pi/U_{BB}$, implying that the odd revival
peaks are damped.

Since the revival period is doubled when $U_{BF}/U_{BB}=(z+0.5)$, all phase factors in
Eq.(\ref{evolution}) equal 1. We conclude that the state revives alternatively to the
anti-symmetric and symmetric coherent states at odd and even peaks, respectively. Here the
anti-symmetric coherent state means
$|\Psi_B\rangle=(\frac{\hat{b}_1^\dag-\hat{b}_2^\dag}{\sqrt2})^N|0\rangle$. To further identify
the difference of the odd-even revival peaks, we examine the average-value of the pseudo-spin $\langle
S_x\rangle$, which is defined by
\begin{equation}
\langle S_x\rangle=\frac{1}{2}\langle\hat{b}_1^\dag\hat{b}_2+\hat{b}_2^\dag\hat{b}_1\rangle.
\end{equation}
Figure 3 displays the temporal evolution of $\langle S_x(\tau)\rangle$. Figure 3(a) explicitly shows that
$\langle S_x\rangle$ alternatively equals $-1$ and $+1$ at the odd and even revival peaks, respectively.
It demonstrates the difference of coherent states at each revival period. In Fig. 3(b), which corresponds
$U_{BF}/U_{BB}=1$, $\langle S_x\rangle$ varies almost synchronously with $\alpha$. No revival peaks are
damped in this case. We suggest an experimental measurement of the physical quantity $\langle S_x\rangle$
to verify our conclusion.

\begin{figure}
\begin{center}
\includegraphics*[width=8.5cm]{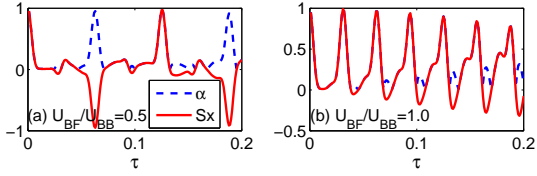}
\caption{(Color online) Temporal evolution of $\langle S_x\rangle$ for (a) $U_{BF}/U_{BB}=0.5$ and (b)
$U_{BF}/U_{BB}=1$. The dashed line is $\alpha(\tau)$.}
\end{center}
\end{figure}

From Eq.(\ref{coherence}), we approximately obtain a linear relation between the damp of the odd revival
peaks and the impurity fillings $\bar m$ as $\alpha_{\textrm{odd peaks}}\approx 1-2\bar m$, providing
that $\alpha_2$ tends to zero at the odd revival peaks. The data points indicate the average of the first three odd revival peaks for $\bar{m}=0.,0.2,0.35,0.5$. The result qualitatively agrees with the
numerical results plotted in Fig. 2(d).

\begin{figure}
\begin{center}
\includegraphics*[width=8.5cm]{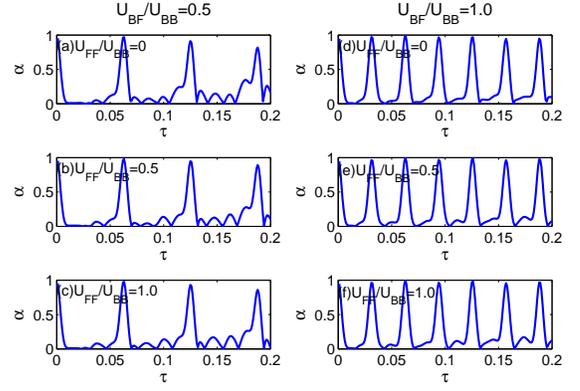}
\caption{(Color online) Temporal evolution of $\alpha$ for $U_{BF}/U_{BB}=0.5$ (left column) and
$U_{BF}/U_{BB}=1$ (right column) for fermion-fermion interactions $U_{FF}/U_{BB}=0.0,0.5,1.0$,
respectively. In all cases, $U_{BB}=100$ and $\bar{m}_\uparrow=\bar{m}_\downarrow=0.5$.}
\end{center}
\end{figure}
\section{Unpolarized fermionic impurities}
We next examine the case of unpolarized fermionic impurities. The Hamiltonian is
\begin{eqnarray}
\hat{H}_2&&=-t_B(\hat{b}_1^\dag \hat{b}_2+\hat{b}_2^\dag\hat{b}_1)+\frac{1}{2}U_{BB}\sum_{i=1,2}\hat{n}_i(\hat{n}_i-1)\nonumber\\
&&-t_F\sum_{\sigma=\uparrow,\downarrow}(\hat{f}_{1\sigma}^\dag \hat{f}_{2\sigma}+\hat{f}_{2\sigma}^\dag
\hat{f}_{1\sigma})+U_{BF}\sum_{i,\sigma}\hat{n}_i\hat{m}_{i\sigma}\nonumber\\
&&+U_{FF}\sum_{i}\hat{m}_{i\uparrow}\hat{m}_{i\downarrow},\label{ham2}
\end{eqnarray}
where $U_{FF}$ is the fermion-fermion interaction and $\hat{m}_{i\uparrow}$ ($\hat{m}_{i\downarrow}$)
are the spin-up (spin-down) fermionic number operators at the $i$-th site. We have chosen
$t_{F_\uparrow}=t_{F_\downarrow}\equiv t_F$ and $U_{BF_\uparrow}=U_{BF_\downarrow}\equiv U_{BF}$. The
average coherence degree is a mixture of the pure bosons, bosons with spin-up or spin-down fermions, and
bosons with both spin-up and spin-down fermions, which is expressed as
\begin{eqnarray}
\alpha=&&(1-2\bar{m}_\uparrow)(1-2\bar{m}_\downarrow)\alpha_1+2\bar{m}_\uparrow(1-2\bar{m}_\downarrow)\alpha_2\nonumber\\
&&+2\bar{m}_\downarrow(1-2\bar{m}_\uparrow)\alpha_3+2\bar{m}_\uparrow\times
2\bar{m}_\downarrow\alpha_4,\label{twofermion}
\end{eqnarray}
where $\alpha_1, \alpha_2,\alpha_3$ and $\alpha_4$ refer to the coherence degree of the pure bosons,
bosons with a spin-up fermion, bosons with a spin-down fermion and bosons with a spin-up and a spin-down
fermions system, respectively. $\bar{m}_\uparrow$ ($\bar{m}_\downarrow$) represent the average fillings
of spin-up (spin-down) fermionic impurities.

Figure 4 show the temporal evolution of coherence degree $\alpha$ for $U_{BF}/U_{BB}=0.5$ (left column)
and $U_{BF}/U_{BB}=1$ (right column), respectively. The coherence collapses and revives as in the case
of the polarized fermionic impurities. Intriguingly, the revivals are irrelevant to the strength of the
fermion-fermion interaction $U_{FF}$. This effect can also be interpreted by the independent evolution
model as described above.

As to the damp of the odd revival peaks, we can deduce a relation with the relative fillings of the
unpolarized fermionic impurities $\gamma=\bar{m}_\downarrow/\bar{m}_\uparrow$ from
Eq.(\ref{twofermion}). Since $\alpha_2=\alpha_3\approx\alpha_4$, we have
\begin{equation}
\alpha\approx(1-2\bar{m}_{eff})\alpha_1+2\bar{m}_{eff}\alpha_2,
\end{equation}
where
$\bar{m}_{eff}=\bar{m}_\uparrow+\bar{m}_\downarrow-2\bar{m}_\uparrow\bar{m}_\downarrow=(1+\gamma)\bar{m}_\uparrow-2\gamma\bar{m}_\uparrow^2$.
We conclude $\alpha_{\textrm{odd peaks}}\approx1-2\bar{m}_{eff}$. Our results may be demonstrated by the
experimental device used by the authors of Ref. [\onlinecite{Will}].

\section{summary}
In summary, we have studied the quantum coherence of a few-boson system with fermionic impurities in a
double-well potential. The damp of the odd revival peaks is closely related to the relative strength of
the boson-fermion and boson-boson interactions. We identify that the symmetric and the anti-symmetric
coherent states appear alternatively at the revival peaks. When the system comprises unpolarized
fermionic impurities, we found that the coherence revivals are irrelevant to the fermion-fermion
interaction.

This work is supported by funds from the Ministry of Science and Technology of China under Grant No.
2012CB821403 and by the NSFC under Grant No. 10874018.

\end{document}